\documentclass[USenglish]{article}
\usepackage[utf8]{inputenc}

\usepackage{graphicx}
\usepackage{amsmath,amssymb,color,comment}

\usepackage{authblk}

\newcommand{\intd}[1]{\int\!\!\diff #1}

\newcommand{\rr}{\mathbf{r}}
\newcommand{\kk}{\mathbf{k}}
\newcommand{\pp}{\mathbf{p}}
\newcommand{\PP}{\mathbf{P}}

\newcommand{\uu}{\mathbf{u}}
\newcommand{\nnu}{\mathbf{\nu}}
\newcommand{\jj}{\mathbf{j}}
\newcommand{\diff}{\mathrm{d}}

\newcommand{\OO}{\mathbf{1}}
\newcommand{\TT}{\mathbf{2}}
\newcommand{\QQ}{\mathbf{q}}\newcommand{\EE}{\mathbf{e}}
\newcommand{\eps}{\varepsilon}
\newcommand{\vv}{\mathbf{v}}
\newcommand{\qq}{\mathbf{q}}
\newcommand{\zero}{\mathbf{0}}
\newcommand{\sflux}{\mathbf{s}}
\newcommand{\bb}{\mathbf{b}}
\newcommand{\nunu}{\mathbf{0}}
\newcommand{\xx}{\mathbf{x}}
\newcommand{\RR}{\mathbf{R}}

\def\re#1{(\ref{#1})}   %% Note: AMSTeX's  \eqref  also does  (\ref{#1})
\def\eqn#1#2{ \begin{align} \label{#1}         #2 \end{align}}

\def\nl#1{          \\ \label{#1}        }  %% newline
\def\nnl#1{ \tag*{} \\ \label{#1}        }  %% nonumber newline
\let\f\frac                     %% fraction styles
\newcommand{\p}{\partial}

\begin{document}

%%%--------------------------------------------%%%
%%% Please do not alter the following 7 lines: %%%
%%%--------------------------------------------%%%
%\articletype{Research Article}
%\journalname{J.~Non-Equilib.~Thermodyn.}
%\journalyear{2015}
%\journalvolume{???}
%\journalissue{???}
%\startpage{1}
%\aop
%\DOI{10.1515/jnet-YYYY-XXXX}
%%%--------------------------------------------%%%

\title{Extra mass flux in fluid mechanics}
%\author{V\'an P.$^{123}$, Pavelka, M.$^{4,5}$, Grmela, M.$^{4}$,   \\
%	$^1$Department of Energy Engineering, BME, Budapest, Hungary and \\
%	$^2$Department of Theoretical Physics, Wigner Research Centre for Physics,\\
%	Institute for Particle and Nuclear Physics, Budapest, Hungary and \\
%	$^3$Montavid Thermodynamic Research Group\\
%	$^4$\'{E}cole Polytechnique de Montr\'{e}al, \\C.P. 6079  succ. Centre-ville, Montr\'{e}al, H3C 3A7,  Qu\'{e}bec, Canada\\
%	$^5$New Technologies - Research Centre, University of West Bohemia,  \\Univerzitn\'{i} 8, 306 14 Pilsen, Czech Republic}

%************ de Gruyter ***************************
%\runningtitle{Extra mass flux}

%\author{V\'an P., Pavelka, M. and Grmela, M.}
\author[1,2,3]{V\'an P.}
\author[4,5,6]{Pavelka, M.}
\author[6]{Grmela, M.}
%\runningauthor{P. V\'{a}n et al.}

\affil[1]{Department of Theoretical Physics, Wigner Research Centre for Physics, Institute for Particle and Nuclear Physics, Budapest, Hungary}
\affil[2]{	Department of Energy Engineering, BME, Budapest, Hungary, e-mail: van.peter@wigner.mta.hu}
\affil[3]{Montavid Thermodynamic Research Group}
\affil[4]{Mathematical Institute, Faculty of Mathematics and Physics, Charles University in Prague, Sokolovska 83, 186 75 Prague, Czech Republic}
\affil[5]{Department of Chemical Engineering, University of Chemistry and Technology Prague, Technicka 5, 16628 Prague 6, Czech Republic}
\affil[6]{\'{E}cole Polytechnique de Montr\'{e}al, C.P. 6079  succ. Centre-ville, Montr\'{e}al, H3C 3A7,  Qu\'{e}bec, Canada}
%***************************************************

\date{\today}

%\abstract{Consequences of   Galilean invariance and   Hamiltonian structure of  fluid mechanics  on mass flux are investigated. The physics behind  the possible appearance of self-diffusion  is discussed in the context of weakly nonlocal extension of classical hydrodynamics.}
%\keywords{keywords}
%\communicated{...}
%\dedication{...}

%\received{...}
%\accepted{...}

\maketitle
%\linenumbers

\begin{abstract}
The conditions of existence of extra mass flux in single component dissipative non-relativistic fluids are clarified. By considering Galilean invariance we show that if total mass flux is equal to total momentum density, then mass, momentum, angular momentum and booster (center-of-mass) are conserved.  However, these conservation laws may be fulfilled also by other means. We show an example of weakly non-local hydrodynamics where the conservation laws are satisfied as well although the total mass flux is different from momentum density. 
\end{abstract}

\section{Introduction}

Mass flux arising in evolution equations of fluid dynamics equals momentum density plus, possibly, other terms. Our objective in this paper is to discuss the other terms from the point of view of Galilean invariance, Hamiltonian structure and kinetic theory. The general discussion is then illustrated in the context of weakly nonlocal  hydrodynamics. Investigations of mass flux have been initiated in
\cite{LanLif59b,DV,Kli92a,Bre06a,Bre12a} and continued more recently  in  \cite{KosLiu98a,Ott07a,Liu2008,HCO-mass-flux,God10a,Miroslav-mass-flux}. Extra mass flux from \cite{DV} is discussed in Appendix \ref{sec.DV}.  

In \cite{HCO-mass-flux} the conclusion was reached that extra mass fluxes are not thermodynamically admissible. We analyze the admissibility requirements used in \cite{HCO-mass-flux} from a different perspective and in a different framework. Galilean invariance in the continuum time-space four-formulation leads to nearly the same conclusion, namely that the difference between mass flux and momentum density should be zero or in a special form such that booster and angular momentum density are conserved. A model of weakly non-local hydrodynamics (beyond the scope of the models considered in \cite{HCO-mass-flux}) is identified to demonstrate the special form. Mass flux thus does not need to be equal to momentum density.

A similar problem is related to the {\em choice of flow-frames} in relativistic fluids \cite{VanBir14p}, where the difference between the particle and energy-momentum flows is more evident. According to the  general consensus the velocity field of a fluid may be chosen conveniently and fixed to the particles (Eckart frame) to the energy (Landau-Lifshitz frame) or in any other reasonable manner. However, the generic instability of dissipative relativistic fluids seems to be related to the choice of flow-frames \cite{KosLiu00a,Osa12a,TsuKun13a,BecEta15a,GarEta15a,VanBir14p}.

In a non-relativistic theory the equations and physical quantities are relative. The {\em reference-frame dependency or independence} is investigated  by transformation rules between the physical quantities, similarly as in \cite{Havas}, in contrast to the covariant physical quantities and governing equations of a relativistic theory (compare them e.g. in \cite{Muller-Ruggeri}). Moreover, the related question of material frame indifference introduces confusing terminology because of the many different formulations and concepts \cite{Fre09a}. Therefore, in this paper we explicitly write whenever a physical quantity or equation is independent of reference frames or independent of flow-frames.  When both properties are fulfilled, then we call the quantity {\em absolute}.

A flow-frame thus characterizes velocity of the fluid. The velocity is, however, still relative as it is different for each inertial observer. Fixing an inertial observer then means choosing a particular reference frame. A physical quantity transforms by changing inertial reference frames by Galilean transformations in accordance with its tensorial properties. Only when both reference and flow-frames are specified, the evolution equations gain a concrete form.

In the following we give the balance equations and thermostatics, and we calculate the entropy production of single component Galilean relativistic fluids following from an absolute theory \cite{Van15m}. We use the well known relative fields, the self-diffusion flux and self-momentum density, which follow from the general treatment. Then flow-frame of the fluid is fixed to the self-momentum field, and we recover the well known form of the substantial balances except the presence of an extra mass flux (interpreted as self-diffusion). The Galilean transformation rules are presented and their application to the balances and the thermodynamic relations show the consistency of the treatment. 

In this paper we show that extra mass flux cannot be eliminated by changing the reference frame or the flow frame of the usual relative quantities of fluid mechanics. Its presence is also consistent with angular momentum conservation. However, the difference of momentum density and mass flux leads to the violation of booster conservation, the requirement of uniform center of mass motion, unless the extra mass flux is in a special form. This way, one can simplify the treatment of \cite{HCO-mass-flux}, clarify the conditions of its appearance and show examples of extra mass flux in various theoretical frameworks. 

Apart from the perspective of Galilean invariance, we also follow the Hamiltonian structure of the equations. An extra velocity-dependent term can be added to the free energy which causes also an extra mass flux and extra terms in the pressure tensor. 

Finally, we show that extra mass flux naturally emerges in non-local kinetic theory.

\section{Extra mass flux}

Three independent arguments, one based on the Galilean invariance (in Section \ref{GI}), the other on Hamiltonian structure (in Section \ref{HS}) and nonlocal kinetic theory (Sec. \ref{sec.KT}), point out to the possibility of a more complex mass flux.

\subsection{Galilean relativistic fluid mechanics: transformation of relative fields, balances}\label{GI}

The starting point of the reference frame independent theory is that the basic densities and their fluxes form a single third order symmetric four-tensor, the mass-momentum-energy density-flux tensor $Z^{\alpha\beta\gamma}$, where $\alpha,\beta,\gamma \in \{0,1,2,3\}$ and $Z^{\alpha\beta\gamma}=Z^{\alpha\gamma\beta}$. In a particular reference frame the tensor can be written in the following form:
\eqn{mme_tens}{
	{\mathbf{Z}}  = 		\begin{pmatrix}			\begin{pmatrix}				\rho & \pp \\ \pp& \EE			\end{pmatrix} &			\begin{pmatrix}				\jj & \PP \\ \PP & \QQ			\end{pmatrix}		\end{pmatrix} \quad
	Z^{\alpha\beta\gamma}  = 
	\begin{pmatrix}
		\begin{pmatrix}
			\rho & p^i \\ p^j & e^{ji}
		\end{pmatrix} &
		\begin{pmatrix}
			j^k & P^{ki} \\ P^{kj} & q^{kji}
		\end{pmatrix}
	\end{pmatrix}
}
Here the mass, momentum and energy densities are $\rho, p^i$ and $e^{ij}$  and the corresponding fluxes are $j^k, P^{ki}$ and $q^{kji}$, where $i,j,k \in \{1,2,3\}$. The energy density and the energy flux are related to the traces of the corresponding tensors $e = e^j_j/2, q^k = q^{kj}_j/2$  \cite{Van15m}. We have used a notation with indexes to indicate the proper tensorial properties and also repeated the formula without indexes to indicate the coordinate free (but reference and flow-frame dependent) nature of the physical quantities. In the following we will use both, whenever the correct interpretation of the tensorial properties require. The Galilean transformation rules of the different relative quantities uniquely follow from the four-vector representation \cite{Van15m}:
\begin{subequations}
\begin{align}
\label{rtr}	\rho' &=  \rho,  \\
\label{jtr}	\jj' &= \jj + \rho \vv, \quad 	
j'^i = j^i + \rho v^i,  \\
\label{ptr}	\pp' &= \pp + \rho \vv, \quad 	
p'^i = p^i + \rho v^i,  \\
\label{etr}	e' &= e + \pp\cdot \vv+\f{\rho}{2}\vv^2, \quad 	
e' = e + p^k v_k + \f{\rho}{2}v^2, \\
\label{Ptr}	\PP' &=  \PP+ \rho \vv\vv + \vv\pp + \jj\vv, \quad 
P'^{ij} =  P^{ij} + \rho v^iv^j + v^ip^k + j^iv^j, \\
\label{qtr}	\qq' &= \qq + \vv\left(e + \pp\cdot \vv + \f{\rho}{2}v^2\right) + \PP\cdot\vv +	\jj \f{v^2}{2}, \\ 
\label{qtri}	&\qquad \qquad  q'^i = q^i + v^i\left(e+ p^k v_k + \f{\rho}{2}v^2\right) + P^{ik}v_k + j^i \f{v^2}{2}.
\end{align}
\end{subequations}
Here  $\vv$ is the relative velocity field of the fluid related to an external inertial observer. The primed quantities at the left hand side are related to an inertial reference frame, those are the observed densities and fluxes, corresponding to the so called 'total' quantities. These are expressed by the co-moving quantities of the local rest frame of the fluid. The traditional 'conductive parts', $\rho \vv\vv$ for the pressure and $e \vv$ for the heat flux, here are parts of the complete Galilean transformation rules. These transformation rules are apparently more complicated than the usual ones for the fluxes $\qq$ and $\PP$. It is remarkable that the transformation formula of the energy corresponds to the relation of  total, kinetic and internal energies when $\pp = \bf 0$. We emphasize again that the transformation rules are rigorously derived in the frame independent formulation \cite{Van15m}. An other way of deriving transformation rules is requiring the Galilean invariance of the complete system of balances \re{sm_rbal}-\re{se_rbal}, considering that the energy balance is the trace of a second order tensorial balance according to the requirements of non-relativistic kinetic theory \cite{Rug89a,Muller-Ruggeri}. 

Frame independent treatment of the basic balances of a single component Galilean relativistic fluid leads to \cite{Van15m}
\eqn{m_bal}{
	\dot \rho+ \rho \nabla\!\cdot\! \vv + \underline{\nabla\cdot \jj} &= 0, \quad
	\dot \rho+ \rho \p_k v^k + \underline{\p_k j^k} = 0, \nl{p_bal}
	\underline{\dot \pp + \pp \nabla\!\cdot\!\vv} + \rho \dot \vv +
	\underline{\jj\! \cdot\!\nabla\vv} + \nabla\! \cdot\! \PP  &= \mathbf{0}, \quad
	\underline{\dot p^i + p^i \p_k v^k} + \rho \dot v^i +
	\underline{j^k \p_k v^i} + \p_k P^{ki}  = 0^i, 	\nl{e_bal}
	\dot e +e \nabla\!\cdot\!\vv +\nabla\!\cdot\!\qq + \underline{\pp\! \cdot\!\dot \vv} +
	\PP\! :\!(\nabla\vv)  &= 0, \quad
	\dot e +e \p_k v^k + \p_k q^k + \underline{p_k \dot v^k} +
	P^{jk}\p_j v_k  = 0.
}

These balance equations are expressed in terms of relative quantities of an inertial observer. We have given them both with and without indexes for the sake of clarity. The first equation is the balance of mass, the second is the balance of momentum and the third is the balance of energy. %The basic fields are the densities of mass, self-momentum and internal energy, $\rho,\pp$ and $e$, respectively. The corresponding conductive current densities (fluxes) are the extra mass flux, the pressure and the heat flux, $\jj,\PP$ and $\qq$. 
The dot denotes the time derivative, $\nabla$ or $\p_i$ is the spatial derivative. The central dot denotes the contraction. The time derivative is best interpreted as substantial time derivative, and the energy corresponds to the usual internal energy.

The underlined nonstandard  terms are related either to the  extra mass flux $\jj$ or to  the extra momentum density $\pp$ (which is not automatically connected to the relative velocity and  therefore not equal to $\rho \vv$ without any further ado). In the above form of the balances the flow-frame is not yet fixed, we have a freedom to fix the fluid velocity to one of the physical quantities. The two basic choices of these two quantities are either the mass-flow, when $\jj= \bf 0$, or the momentum-flow, when $\pp=\bf 0 $. 

Let us choose the momentum-flow. We obtain therefore the following balances
\eqn{sm_rbal}{
	\dot \rho+ \rho \nabla\cdot \vv + {\nabla\cdot\jj} &= 0, 
	\quad \dot \rho+ \rho \p_k v^k + \underline{\p_k j^k} = 0, \nl{sp_rbal}
	\rho \dot \vv + {\jj\cdot\nabla \vv} + \nabla\cdot\PP &={\mathbf 0}, \quad \rho \dot v^i +
	\underline{j^k \p_k v^i} + \p_k P^{ik}  = 0^i, 	 \nl{se_rbal}
	\dot e +e \nabla\cdot\vv +\nabla\cdot \qq + \PP :(\nabla \vv) &= 0,  \quad
	\dot e +e \p_k v^k + \p_k q^k + 
	P^{jk}\p_j v_k  = 0.
}
These equations have  two interesting features. Firstly, we can see that fixing the momentum-flow does not eliminate the derivative of the relative velocity from the momentum balance. That term is an inertial term related to the apparent change of the relative momentum for the external inertial reference frame. Secondly, the extra mass flux $\jj$ or the self-momentum density, $\pp$, can be eliminated by fixing the flow. The extra mass flux  remains  when the extra momentum (self-momentum) is chosen to be zero.

The Galilean transformation rules of the time-and-space derivatives come from the fact that the four-derivative is a four-covector. The rules themselves are well known: the space derivative is Galilean-invariant $\nabla' = \nabla$, and the transformation rule of the time derivative is the local-substantial relation: $\f{d'}{dt} = \f{d}{dt} - \vv\nabla$, where  $\f{d}{dt}$  denotes the over-dot of the previous equations and the transformed derivative (with prime) is usually denoted by $\partial_t$.

The 'local' form of the basic balances is obtained by applying the above transformation rules to the 'substantial' balances \re{sm_rbal}, \re{sp_rbal} and \re{se_rbal}:
\eqn{tm_fbal}{
	\partial_t \rho +  {\nabla\cdot\jj'} &= 0, \quad
	\partial_t \rho +  \p_k j'^k = 0,\nl{tp_fbal}
	\partial_t (\rho\vv) + \nabla\cdot \PP' &= \zero, \quad
	\partial_t (\rho v^i) + \p_k P'^{ki} = 0^i, \nl{te_fbal}
	\partial_t e' + \nabla\cdot \qq'  &= 0, \quad
	\partial_t e' + \p_k q'^k  = 0.
}
The energy balance \re{te_fbal} is the usual balance of the total energy as it is seen by using transformation rules \re{etr} and \re{qtr}. The total momentum density appears only with its 'conductive' part with the momentum due to the relative motion because $\pp = \zero$.

Finally, the substantial balance of entropy is
\eqn{sm_sbal}{
	\dot s+ s \nabla\cdot \vv + {\nabla\cdot\sflux} = \Sigma \geq 0,
}
where $s$ is the entropy density and $\sflux$ is the entropy flux. The entropy density is Galilean invariant, the entropy flux transforms according to vector Galilean transformation
\eqn{str}{
	s' =  s,  \qquad \sflux'= \sflux + s \vv.
}
With the help of these transformation rules the local balance of entropy is obtained according to our expectations:
\eqn{ls_bal}{
	\partial_t s' +  {\nabla \sflux'} = \Sigma' \geq 0,
}
The entropy production is the measure of dissipation, therefore it is expected to be absolute and therefore also Galilean invariant:
\eqn{ep_eq}{
	\Sigma = \Sigma' \geq 0.
}

\subsubsection{Angular momentum and booster}
	
At the beginning  we have assumed a particular symmetry of the mass-momentum-energy density-flux tensor in order obtain well-interpretable relations. This symmetry has nothing to do with the conservation of angular momentum. For example, the symmetry of the relative pressure does not follow from this assumption as can be seen in the reference-frame dependent form of the tensor given in \re{mme_tens}. 
	
The simplest way of introducing the angular momentum conservation is to restrict ourselves to the mass-momentum density-flux tensor, which can be written in a particular reference frame as:
\eqn{mm_tens}{
	{\mathbf{z}}  = 	\begin{pmatrix}		\rho & \pp \\ \jj & \PP 	\end{pmatrix} \quad
		z^{\alpha\beta}  = 
		\begin{pmatrix}
			\rho & p^i \\ j^k & P^{ki} 
	\end{pmatrix}.
}
	
	In this case the four-angular momentum is the booster and the usual three-angular momentum together, $\Phi^{\alpha\beta\gamma} = x^\alpha z^{\beta\gamma} - x^\gamma z^{\beta\alpha} $,
	\eqn{am_tens}{
		{\mathbf{\Phi}}  = 	\begin{pmatrix}	t\\ \xx \end{pmatrix} \wedge 	\begin{pmatrix}		\rho & \pp \\ \jj & \PP 	\end{pmatrix} = 
		\begin{pmatrix}
			\begin{pmatrix}
				0 &  t\pp- \rho\xx \\ \nunu & t \PP - \jj\xx 
			\end{pmatrix}
			\begin{pmatrix}
				-t\pp+ \rho\xx  & \pp\xx -\xx\pp \\  -t \PP + \jj\xx  & \xx\wedge \PP  
			\end{pmatrix}	
		\end{pmatrix} \nnl{hm}
		\Phi^{\alpha\beta\gamma}= x^{[ \alpha} z^{\beta\gamma ]}=
		\begin{pmatrix}
			\begin{pmatrix}
				0 &  t p^i- \rho x^i \\ 0^j & t P^{ji} - j^j x^i 
			\end{pmatrix}	
			\begin{pmatrix}
				\rho x^k -t p^k & p^i s^k - p^k x^i \\  j^j x^k -t P^{jk}  & x^kP^{ji} -   x^iP^{jk} 
			\end{pmatrix}
		\end{pmatrix} 
	}
One can see that conservation of the four-angular momentum in a particular reference frame requires conservation of booster, see Appendix \ref{sec.booster} or \cite{KosLiu98a}, density and flux of which are $t\pp- \rho\xx$ and $t \PP - \jj\xx $, respectively, together with angular momentum, density and flux of which are $\pp\xx -\xx\pp $ and $\xx\wedge \PP$, respectively. Here the external product is understood in the second order of the pressure, indicated also above, with the help of the index notation. 
	
In the following we show a simplified calculation of the booster-angular momentum conservation for fluids. Our starting point -- the basic assumption -- is analogous to the special relativistic four-angular momentum conservation. We assume that the antisymmetric part of the momentum of the divergence of mass-momentum density-flux tensor is zero for all reference points. In a four-dimensional notation ($\alpha,\beta, \gamma = (0,1,2,3)$ ) it can be written as
\eqn{4consmm}{
	2 x^{[\gamma } \partial_\alpha z^{\alpha\beta]} = x^{\gamma } \partial_\alpha z^{\alpha\beta} -x^{\beta } \partial_\alpha z^{\alpha\gamma}  = \partial_\alpha (x^{\gamma } z^{\alpha\beta} -x^{\beta } z^{\alpha\gamma} ) - z^{\gamma\beta}+z^{\beta\gamma} = 0
}
Therefore, the requirement of conservation of four-angular momentum $\Phi^{\alpha\beta\gamma} = \f{1}{2}\left(x^{\gamma } z^{\alpha\beta} -x^{\beta } z^{\alpha\gamma}\right)$ leads to the symmetry of the mass-momentum.
		
Introducing momentum flow and using the reference frame of the primed quantities \re{rtr}-\re{qtri}, that is an intertial laboratory frame where the balances are  \re{tm_fbal}-\re{te_fbal}, leads to the following relative form of the formulas above:
\begin{gather*}
		\begin{pmatrix} t\\ x^k 	\end{pmatrix}
		\begin{pmatrix}
		\partial_t \rho +\partial_j j'^j  & \partial_t \rho v^i +\partial_j P'^{ji} 
		\end{pmatrix} -	
		\begin{pmatrix}
		\partial_t \rho +\partial_j j'^j  \\ \partial_t \rho v^k +\partial_j P'^{jk} 
		\end{pmatrix}
		\begin{pmatrix}	t& x^i 	\end{pmatrix}  = \nnl{b2}
		\begin{pmatrix}
		0 &  t(\partial_t \rho v^i +\partial_j P'^{ji}) -x^i( \partial_t \rho +\partial_j j'^j ) \\ 
		-x^k( \partial_t \rho +\partial_j j'^j )- t(\partial_t \rho v^k\!\! +\partial_j P'^{jk})\!\!\!\! & 
		x^k(\partial_t \rho v^i +\partial_j P'^{ji} ) - x^i(\partial_t \rho v^k +\partial_j P'^{jk} ) 
		\end{pmatrix}  \nnl{b3}
		=\begin{pmatrix}
		0 & 0^i \\  0^k  & 0^{ki}
		\end{pmatrix},
\end{gather*}
and we obtain 
\begin{subequations}\label{bam_bal}
 \begin{eqnarray}
\label{b_bal}\partial_t(t \rho v^i - x^i \rho ) - \rho v^i + \partial_j(tP'^{ji} - x^ij'^j) + j'^i &=& 0, \\
\label{am_bal}\partial_t( \rho v^i x^k - \rho x^i v^k ) + \partial_j(x^kP'^{ji} - x^iP'^{jk}) -P'^{ki} + P'^{ik} &=& 0.
 \end{eqnarray}
\end{subequations}
As a consequence, one can see that the conservation of the booster, with a density and flux given by $(t \rho v^i - x^i \rho \mbox{ and } tP'^{ji} - x^ij'^j)$ requires that $\rho v^i=j'^i = \rho v^i + j^i$, leading to $j^i=0^i$ or $\jj$ in a special form, namely being divergence of a tensor field so that Eq. \eqref{b_bal} is fulfilled in its integral form. Moreover, conservation of the angular momentum is satisfied when the pressure tensor is symmetric or in a special form such that Eq. \eqref{am_bal} is satisfied in its integral form, see Sec. \ref{Ill} for an example of the special form.
	
%The first condition requires particular attention. Eq. \eqref{b_bal} is the expression of the center of mass conservation in a continuum framework. One might conclude that without internal angular momentum or internal booster (given by the respective expressions involving $\jj$ and $\pp$) the usual physical quantities and balances of simple fluids are recovered because then $\pp = \nunu$ and $\jj = \nunu$ at the same time. On the other hand, a source term in the balance of internal booster may counterbalance the non-zero extra mass flux although still preserving the total booster conservation. In particular, an extra mass flux may have a divergence form, $j^i = \partial_j \psi^{ij}$, when the the booster is conserved. Then the identification of the extra mass flux requires the separation of the true booster flux under the divergence. We will see a particular example in the following.
	
The above train of thoughts is analogous to the usual special relativistic calculations (see e.g. \cite{Heh76a}). However, in a Galilean relativistic framework our third-order basic quantity enables several broader generalizations that can be explored. Note also that the role of relativistic four-spin, which plays the role of internal four-angular momentum, is not completely understood, especially when flow-frame independence is expected \cite{BohVig58a,Heh76a,Bec11a,BroCap02a}. In particular, the non-relativistic counterpart of the usual Frenkel condition eliminates the time-like component of the internal angular-momentum, which eliminates the possibility of an internal booster.

\subsection{Hamiltonian structure}\label{HS}

Clebsch in \cite{Clebsch} and later Arnold in \cite{Arnold}  recognized  Hamiltonian structure in the Euler fluid mechanics equations. In order to see the structure, we have to replace the state variables
$(\rho,\vv,e)$ used in the previous section with $(\rho,\uu,s)$. The mass density $\rho$ remains the same, $\uu$ is the field of relative momentum of an inertial observer ($\uu=\pp'$ of the previous section) and $s$ is the entropy field. The transformation $(\rho,\vv,e)\rightarrow(\rho,\uu,s)$, given by the  relations
\begin{equation}\label{tr}
\rho=\rho;\,\,\uu=\rho\vv;\,\,s=s(\rho,e)
\end{equation}
is one-to-one (since $s\rightarrow e$ is one-to-one because  $\partial s/\partial e$, having the physical meaning of inverse of the absolute temperature, is always positive). The total energy $E$ is given by
\begin{eqnarray}\label{energy}
&&E=E_{kin}+E_{exk}+E_{int}\nonumber \\
&&E_{kin}=\int d\xx \frac{\uu^2}{2\rho};\,\,E_{exk}=\int d\xx e_{exk}(\nabla\vv, \rho,s);\,\,E_{int}=\int d\xx e(\rho,s)\nonumber \\
\end{eqnarray}
where $\xx$ denotes the position vector, $e$ is the local internal energy and $E_{exk}$ is a contribution to the kinetic energy due to  a finer scale motion of the continuum. For example in Section \ref{Ill}, where we consider highly compressible fluids with large gradients of mass density and velocity (fluids in the vicinity of gas-liquid phase transition), the extra kinetic energy $e_{exk}\sim (\nabla\cdot\vv)^2$ (motivated by \cite{Miroslav-JSP2008}) is the kinetic energy of the volume (and mass) changes of  fluid particles (note that $\nabla\cdot\vv$ is a velocity of the volume change).

Following Arnold \cite{Arnold},  the Euler  part of the fluid mechanics equations (governing the nondissipative and  reversible time evolution)  can be cast into the form
\begin{equation}\label{Ham}
\frac{\partial}{\partial t}\left(\begin{array}{ccc}\rho\\ \uu\\s\end{array}\right)=L\left(\begin{array}{ccc}E_{\rho}\\ E_{\uu}\\E_s\end{array}\right)
\end{equation}
that manifestly displays the Hamiltonian structure. We indeed see in (\ref{Ham}) that the gradient of energy
$(E_{\rho},E_{\uu},E_s)$ (that is  a covector)  is transformed into vector by the operator  $L$ called a Poisson operator. The operator  $L$ is given by the Poisson bracket
\begin{eqnarray}\label{bracket}
\{A,B\}&=& \int d\rr (A_{\rho},A_{\uu},A_s)L(B_{\rho},B_{\uu},B_s)^T\nonumber \\
&=& \int d\rr \left[\rho((\partial_i A_\rho) B_{u_i} - (\partial_i B_\rho) A_{u_i}) + u_i((\partial_j A_{u_i}) B_{u_j} - (\partial_j B_{u_i}) A_{u_j})\right.\nonumber \\
&&\left.+s((\partial_i) A_s B_{u_i} - (\partial_i B_s) A_{u_i})\right]
\end{eqnarray}
where $A$ and $B$ are sufficiently regular functions of $(\rho,\uu,s)$; we use the shorthand notation $A_{\rho(\rr)}=\frac{\delta A}{\delta \rho(\rr)}$, where $\delta/\delta$ denotes an appropriate functional derivative.  The easiest way to obtain the explicit form of Eqs.(\ref{Ham}) is to write (\ref{Ham}) as $\frac{dA}{dt}=\{A,E\}$ holds for all $A$ (i.e. $\frac{dA}{dt}=\int d\rr \left[A_{\rho}\frac{\partial\rho}{\partial t}\\+A_{\uu}\frac{\partial\uu}{\partial t}+A_{s}\frac{\partial s}{\partial t}\right]=\int d\rr\left[A_{\rho}(\bullet)+A_{\uu}(\bullet\bullet)+A_{s}(\bullet\bullet\bullet)\right]$. Here the terms represented  by the symbols $(\bullet),(\bullet\bullet),(\bullet\bullet\bullet)$ are obtained by rewriting $\{A,E\}$ with the use of integrations by parts in which the boundary conditions are assumed to be such that all the integrals over the boundary that arise in the calculations equal zero. The time evolution equations are thus  $\frac{\partial\rho}{\partial t}=(\bullet);\, \frac{\partial\uu}{\partial t}=(\bullet\bullet);\,\frac{\partial s}{\partial t}=(\bullet\bullet\bullet)$). Specifically, we arrive at $(\bullet)=-\partial_i(\rho E_{u_i})=-\partial_i\left(u_i+\rho \delta E_{exk}/\delta{u_i}\right)$, which then means that
\begin{equation}\label{jj}
%j_i=-\partial_l\left((E_{exk})_{\partial_l u_i}\right)
j_i=\rho \frac{\delta E_{exk}}{\delta u_i}.
\end{equation}

\subsection{Kinetic theory}\label{sec.KT}
Evolution of fluids is often described within kinetic theory, with the help of the Boltzmann equation. The irreversibilities stem from the collision integral, which is typically strictly local, i.e. depends only on distribution function at a particular point in real space, see e.g. \cite{dGM,Lib90b,MulRug98b}. However, the collision integral can be also completely (or weakly) nonlocal as in \cite{Grmela1997}, or in Enskog kinetic equation \cite{Enskog}. Non-locality of the collision integral means that the forces acting on colliding particles extend over non-negligible area or that the particles have finite size.
\begin{multline}\label{eq.Boltzmann}
\frac{\partial f(\rr_1,\pp^1,t)}{\partial t} = -\frac{\partial}{\partial r_1^k}\left(f \frac{\partial E_f}{\partial p^1_k}\right) +\frac{\partial}{\partial p^1_k}\left(f \frac{\partial E_f}{\partial r_1^k}\right) +\\
+\intd\OO'\intd\TT\intd\TT' W(f;\OO,\OO',\TT,\TT') (f(\OO)'f(\TT)'-f(\OO)f(\TT)),
\end{multline}
where the collision integral is constructed from a collision kernel $W$. In the case of standard Boltzmann collision integral, the kernel is zero if positions of particles $\OO$, $\OO'$, $\TT$, $\TT'$ are not the same.

Mass has to be conserved in the collisions regardless non-locality of the collision integral. In the standard Boltzmann collision integral, which is strictly local in space, mass is conserved at each point of space simply because the distribution function has the same normalization (integral over phase space) before and after the collision. On the other hand, in the case of generally nonlocal collisions, where the collision kernel also depends on space, conservation of mass is a requirement on the form of the collision kernel such that phase-space integral the right hand side of Eq. \eqref{eq.Boltzmann} is zero. In other words, integral with respect to $\pp^1$ of the collision integral is a divergence in space of a vector field, 
\begin{equation}
\intd\pp^1\intd\OO'\intd\TT\intd\TT' W(f;\OO,\OO',\TT,\TT') (f(\OO')f(\TT')-f(\OO)f(\TT)) = -\partial_i j_i,
\end{equation}
the field being the extra mass flux.

It is thus natural in nonlocal kinetic theory that there is an irreversible extra mass flux in the evolution equation of density.

\section{Constitutive relations}\label{CR}

Having established the  possible  appearance of the extra mass flux $\jj$, we now proceed to specify it. A function expressing $\jj$ in terms of the  fields that form the set of independent state variables is called a constitutive relation.
We take two routes to discuss it. Both routes need to supplement  equations governing the time evolution of the fields chosen to characterized states with another equation governing the time evolution of  the entropy field  $s$ used already in  Section \ref{HS}. We recall that
$s$ is not an independent field but a field that depends on the  fields included into the set of independent state variables.  On both routes the specification of the constitutive relations
is based on the requirement that the entropy produced  during the time evolution is non-negative. On the first route (in Section \ref{DS})   this requirement is directly employed.  On the second route (in Sections \ref{IS} and \ref{Ill}) we first  adopt a field closely related to $\jj$  as an independent state variable with its own time evolution equation (that is coupled to the equations governing the time evolution of the fields of mass, momentum, and energy). From the requirement of non-negativity of the entropy production and from an another requirement addressing the relative speed  of the relaxation of $\jj$ and of the remaining fields we then arrive at the constitutive relation for $\jj$. On both routes we show that self-diffusion is a possible constitutive relation for $\jj$.

\subsection{Direct specification}\label{DS}

The usual Gibbs relation
\eqn{klGibbs}{
	de = Tds + \mu d\rho,
}
holds for the fields $e,s,\rho$;  $T$ is the absolute temperature and $\mu$ is the chemical potential. The extensivity condition and the form of the entropy current density are obtained from the relation between the corresponding four quantities \cite{Van15m}:
\eqn{saramfb}{
	e+p = Ts +\mu\rho, \qquad 	\sflux = \f{1}{T} \left(\qq -\mu \jj\right),
	\quad s^i = \f{1}{T} \left(q^i -\mu j^i\right),
}
$p$ is the static pressure. The given entropy flux is straightforward also considering the nonzero mass flux. The transformation rules of the temperature, the pressure, and the chemical potential are due to the four-cotensor form of the intensive quantities and the momentum-flow:
\eqn{ttr}{
	T' = T, \qquad p' =p, \qquad \mu'  = \mu - \f{v^2}{2}.
}

These transformation rules result in the following form of the transformed Gibbs-relation and the extensivity condition:
\eqn{Gibbs_frel}{
	de' = Tds +\mu' d\rho + \vv\cdot d(\rho \vv), \qquad 	e'+p = Ts +\mu'\rho + \rho v^2
}
One can see that for the external observer the relative momentum density, $\rho \vv$, is an extensive quantity and that the related intensive is the relative velocity $\vv$. 

Now we proceed to identify a possible constitutive relation for the extra mass flux $\jj$.
The entropy production can be calculated by using the substantial balances of mass and internal energy \re{sm_rbal} and \re{se_rbal} together with the Gibbs relation \eqref{klGibbs} and entropy current density \re{saramfb}:
\eqn{eprod}{
	\dot s(e,\rho) + s \nabla\cdot \vv + \nabla\cdot \sflux= \f{1}{T} \dot e -\f{\mu}{T} \dot \rho +s \nabla\cdot\vv + \nabla\cdot\left(\f{\qq -\mu \jj}{T} \right) \nnl{eprod1}
	=-\jj \cdot\nabla \f{\mu}{T}+
	\qq\cdot\nabla\f{1}{T} -
	\f{1}{T}\left(\PP -p{\mathbb I}\right):(\nabla\vv) =\Sigma \geq 0.
}

Here ${\mathbb I}$ is the unit tensor. For isotropic fluids one obtains the following linear relations for the thermodynamic fluxes and forces:	
\begin{subequations}\label{diff_c}
\begin{eqnarray}
	\label{jjc}\jj &=& -\xi \nabla\f{\mu}{T} + \chi_1 \nabla\frac{1}{T},\\
	\label{hvez_c}\qq &=& -\chi_2 \nabla\f{\mu}{T} + \lambda \nabla\frac{1}{T},\\
	\label{viszk_c}\PP &=& p{\mathbb I} - \eta_v \nabla\cdot \vv {\mathbb I}-
	\eta \left(\nabla\vv+\vv\nabla - \f{2}{3}\nabla\cdot\vv {\mathbb I} \right).
\end{eqnarray}
\end{subequations}
%\eqn{diff_ci}{
%	j^i &= -\xi \p^i\f{\mu}{T} + \chi_1\p^i\frac{1}{T},\nl{hvez_ci}
%	q^i &= -\chi_2 \p^i\f{\mu}{T} + \lambda \p^i\frac{1}{T},\nl{viszk_ci}
%	P^{ij} &= p\delta^{ij} - \eta_v \p_kv^k \delta^{ij}-
%	\eta \left( \p^iv^j+ \p^jv^i - \f{2}{3} \p_kv^k\delta^{ij} \right).
%}
The first term on the right hand side of the first line represents indeed self-diffusion. Here Onsager symmetry \cite{dGM} requires that $\chi_1=\chi_2$.  Moreover, $\eta\geq 0$, $\eta_v\geq 0$, $\xi\geq 0$, $\lambda\geq 0$ and $\lambda\xi-(\chi_1+\chi_2)^2/4\geq 0$ because of the second law (the non-negativity of the entropy production).
Considering parity with respect to time-reversal \cite{PRE2014}, one can conclude that extra mass flux \eqref{jjc} causes dissipative irreversible evolution.

It is interesting but not apparent from this form that the entropy production is invariant under Galilean transformations. Moreover, every thermodynamic flux and force is invariant, too. 

%To fulfill booster and angular momentum conservations \eqref{bam_bal}, coefficients $\xi$ and $\chi$ are necessarily constant and self-momentum $\pp$ is equal to $\jj$, which makes the total mass flux equal to the total momentum density. See Sec. \ref{Ill} for an example where total mass flux and total momentum density are not the same.

%Considering the results of Section 2.1.1 about booster conservation, one can conclude that that extra mass flux without internal booster is dissipative. 

\subsection{Indirect specification in extended fluid mechanics}\label{IS}

As we have seen in Section \ref{HS}, the extra mass flux $\jj$  arises due to an extra contribution $E_{exk}$ to the kinetic energy (see Eq. \eqref{energy}). We now make an extension of the formulation presented in Section \ref{HS}. The extension is made in four steps.

\textit{Step 1}:
We replace  $(\rho,\uu,s)$ with $(\rho,\uu,s,\kk)$, where $\kk$ is a field of odd parity (i.e. in changes sign if the sign of time is changed; $\uu$ has odd parity and $\rho$ and $s$ have even parity - see more \cite{PRE2014}). Next, we specify the energy $E(\rho,\uu,s,\kk)$.

\textit{Step 2}:
We  replace the Poisson bracket (\ref{bracket}) with a new Poisson bracket involving also $\kk$. We require that the extended bracket reduces to the bracket (\ref{bracket}) if $\kk$ is absent.

\textit{Step 3}:
The equation governing the time evolution of $\kk$ (the equation is obtained in Step 2) is supplemented with a term that: (i) makes to relax to  $0$  (i.e. $\kk\rightarrow 0$)  as $t\rightarrow\infty$ and (ii) makes the entropy to grow in the time evolution.

\textit{Step 4}:
The relaxation of $\kk$ is fast and in its last stage  $\kk$ becomes enslaved to the remaining fields (i.e. $\kk$ becomes a function of the remaining fields). If this enslaved $\kk$ is then inserted into the extra mass flux $\jj=\rho E_{\uu}-\uu$ a self-diffusion contribution to the mass flux arises.

This type of extension, where the evolution of the extra state variable is given by the extended Poisson bracket, has to be seen as reversible evolution. Indeed, the Poisson bracket only provides reversible evolution, see \cite{PRE2014}. On the other hand, the extra mass flux gives irreversible evolution. The irreversibility appears in steps 3 and 4, where the extra state variables becomes enslaved by the slower variables. It means that one can either prescribe a constitutive relation (the direct approach) for the extra variable and the extra variable then provides irreversible evolution right away, or one can at first extend the Poisson bracket to obtain an evolution equation for the extra variable, which relaxes quickly to a value given by the slower variables. After the relaxation, the dependence of the extra variable on the slower variables serves as the constitutive relation, and the evolution caused by the extra state variable is no longer reversible. This shows the physical origin of constitutive relations from Sec. \ref{DS}.

This type of extension in which self-diffusion arises has been made in \cite{Miroslav-mass-flux} with $\kk=$ {\em one particle distribution function} $f(\xx,\nnu)$ that provides an extra fine-scale information about fluids. The fine-scale velocity contributing  to the kinetic energy is $\nnu$. In the following section we shall illustrate the extension leading to self-diffusion in the setting that we have started in Section \ref{HS}.

An another possibility is to introduce the fine-scale details into the energy or free energy functional. This route is followed for example in the Cahn-Hilliard model \cite{CH}, where non-local terms are added to the free energy functional. Physical motivation for such extension of free energy is the same as in the preceding paragraphs. An extra state variable (e.g. an extra mass flux or fluid particle volume) naturally contributes to the free energy as for example in the Extended Irreversible Thermodynamics \cite{Jou-EIT}. When the extra state variable becomes enslaved by the remaining (hydrodynamic) variables, its contribution to the free energy does not disappear, and it is represented by the non-local terms given by the enslaved extra state variable. This is the way how non-locality is introduced in Sec. \ref{Ill}. Note that due the presence of the extra terms in the free energy functional, the extra evolution caused by the terms can be both reversible or irreversible.

\section{Illustration in weakly nonlocal hydrodynamics}\label{Ill}
\subsection{Free energy}
In this illustration  we make the extension of the setting introduced in   Section \ref{HS} by considering  $\nabla\cdot\vv$ as the enslaved extra  field   $\kk$. From the physical point of view, $\nabla\cdot\vv$ is the velocity of  changes of the volume of the fluid particle. For the sake of simplicity we limit ourselves in this illustration  to isothermal fluids. This means that the temperature $T$ is a constant and   we can omit the entropy $s$ from the state variables. Moreover, the energy $E$ in (\ref{Ham}) has to be replaced by the free energy that we denote by the symbol $\Phi$. The first step in the extension is thus $(\rho,\uu,s)\rightarrow (\rho,\uu,\nabla\uu)$. We choose the free energy to be
\begin{equation}\label{eq.Phi}
\Phi = \int d\xx \left[\frac{\uu^2}{2\rho} +  \frac{1}{2}\rho\sigma \left(\nabla\cdot \frac{\uu}{\rho}\right)^2 + \varphi(\rho,T)\right],
\end{equation}
where the constant parameter  $\sigma$ is related to the  surface area of the fluid particles. Physical consequences of this free energy are discussed in Appendix \ref{sec.phys.conseq}.

\subsection{Hamiltonian evolution}
Now we proceed to the second step. Since in this illustration we are not adopting any new field as an independent state variable, the bracket (\ref{bracket}) remains unchanged, except that the functional derivatives appearing in it become variational derivatives (because we  allow $A$ and $B$ depend also on the spatial gradients of the state variables) and, moreover, the last term in the bracket is missing since we are omitting the field $s$ from the set of independent state variables (due to our limitation to isothermal systems). Another difference, again due to the limitation to isothermal system, is that the energy $E$ in (\ref{Ham}) is replaced by the free energy $\Phi$ (that is given in (\ref{eq.Phi})).

Reversible evolution of a functional $A$ of the state variables is then given by
\begin{equation}
\dot{A} = \{A, \Phi\} = \int d\xx A_\rho \frac{\partial \rho}{\partial t} + A_{u_i} \frac{\partial u_i}{\partial t}, 
\end{equation}
which then leads to reversible evolution equations
\begin{subequations}
\begin{eqnarray}
\left(\frac{\partial \rho}{\partial t}\right)_{\mbox{rev}} &=& -\partial_i(\rho \Phi_{u_i})\\
\left(\frac{\partial u_i}{\partial t}\right)_{\mbox{rev}} &=& -\rho\partial_i \Phi_\rho - \partial_j(u_i \Phi_{u_j}) - u_j\partial_i \Phi_{u_j}.
\end{eqnarray}
\end{subequations}
Using Eq. \eqref{eq.Phi} leads to
\begin{subequations}
	\begin{eqnarray}
	\label{eq.evo.rho}\left(\frac{\partial \rho}{\partial t}\right)_{\mbox{rev}} &=& -\partial_i u_i + \partial_i \partial_i (\rho \sigma  \partial_j v_j),\\
	\label{eq.evo.ui}\left(\frac{\partial u_i}{\partial t}\right)_{\mbox{rev}} &=& -\partial_i (\rho \varphi_\rho - \varphi) - \partial_j \left(\frac{u_i u_j}{\rho}\right) \nonumber\\
	&&\underbrace{- \partial_j \left(\partial_i v_j \sigma \rho \partial_k v_k - v_i \partial_j(\sigma \rho \partial_k v_k)\right)}_{\mbox{higher-order terms}},
	\end{eqnarray}
	where identity $\partial_i(\rho\varphi_{\rho} - \varphi) = \rho\partial_i \varphi_\rho$ was used. Velocity was defined as $\vv = \uu/\rho$. Note that we do not interpret this velocity as an average velocity of particles constituting the fluid. Average velocity of particles should be rather $\bar{\vv}=\Phi_\uu$ because that is the velocity in the Lie algebra semidirect product leading the the hydrodynamic Poisson bracket, see \cite{Marsden-Ratiu-Weinstein}.
\end{subequations}
From Eq. \eqref{eq.evo.rho} it follows that the extra mass flux is given by 
\begin{equation}\label{eq.j}
 \jj = -\nabla(\rho \sigma \nabla \cdot (\uu/\rho)).
\end{equation}

\subsection{Irreversible evolution}
In order to introduce relaxation of $\vv$ toward spatial homogeneity, we introduce the standard Navier-Stokes-like volumetric dissipation
\begin{equation}
-\frac{\delta}{\delta \Phi_{u_i}}\underbrace{\intd\xx \frac{1}{2}\eta_{vol}\rho (\nabla\cdot\Phi_\uu)^2}_{\mbox{dissipation potential}} = \partial_i\left(\eta_{vol} \rho \partial_k \Phi_{u_k}\right)
\end{equation}
into the equation for momentum, Eq. \eqref{eq.evo.ui}, which thus becomes
\begin{eqnarray} 
\label{eq.evo.ui.full}\frac{\partial u_i}{\partial t} &=& -\partial_i (\rho \varphi_\rho - \varphi) - \partial_j \left(\frac{u_i u_j}{\rho}\right) \nonumber\\
&&- \partial_j \left(\partial_i v_j \sigma \rho \partial_k v_k - v_i \partial_j(\sigma \rho \partial_k v_k)\right)\nonumber\\
&& \underbrace{+\partial_i\left(\eta_{vol}\rho \partial_j\left(\frac{u_j}{\rho}-\frac{1}{\rho}\partial_j \left(\sigma\rho\partial_k \frac{u_k}{\rho}\right)\right)  \right)}_{\mbox{irreversible terms}},
\end{eqnarray} 
where $\eta$ is the kinematic viscosity coefficient and constant $\eta_{vol}$ is proportional the volume viscosity coefficient. Note that we do not pay attention the upper or lower positions of the indexes as we consider the Cartesian metric. Evolution equation for density, \eqref{eq.evo.rho}, remains the same. Equations \eqref{eq.evo.rho} and \eqref{eq.evo.ui.full} represent the evolution equations of the weakly non-local fluid.

\subsection{Admissibility criteria}
Let us now show that all admissibility criteria required in \cite{HCO-mass-flux} are fulfilled for evolution equations \eqref{eq.evo.rho} and \eqref{eq.evo.ui.full}.

Total mass and momentum are clearly conserved. Total angular momentum,
\begin{equation}
\intd\rr \eps_{ijk}x^j u^k,
\end{equation}
is also conserved as can be verified by straightforward calculation. Pressure can be introduced as
\begin{equation}
p = \rho \varphi_\rho - \varphi
\end{equation}
and thus the first term in the momentum equation can be seen as $-\partial_i p$. We assume that both $\eta_{vol}$ and $\sigma$ are constant. Let us also denote the right hand side of Eq. \eqref{eq.evo.ui.full} as divergence of the total pressure tensor $\mathbf{P}'$ although such a notation might not be fully compatible with \eqref{Ptr} due to the extra term in energy. 

An additional conservation law was suggested in \cite{KosLiu98a} or on p. 196 of \cite{LanLif59b}, namely the booster density
\begin{equation}
\bb = \rho \xx - \uu t
\end{equation}
conservation, which was also required in \cite{HCO-mass-flux}. Physical meaning of booster conservation is explained in Appendix \ref{sec.booster}.
Taking equations \eqref{eq.evo.rho} and \eqref{eq.evo.ui.full}, we obtain that
\begin{eqnarray}\label{Ham.b}
\frac{\partial b^i}{\partial t} &=& \frac{\partial \rho}{\partial t}x^i - u^i   
-t \frac{\partial u^i}{\partial t}
\nonumber\\
&=& -(\partial_j u^j) x_i  - (\partial_j j^j) x^i - u^i + \partial_j(t P'^{ij})\nonumber\\
&=& \partial_j(-x^i u^j - x^i j^j) + j^i + \partial_j(t P'^{ij}),
\end{eqnarray}
and booster is thus locally conserved because the extra mass flux is also divergence of a tensor field, see Eq. \eqref{eq.j}. This balance equation has the same structure as Eq. \eqref{b_bal}, with the source term $\jj$ not really playing the role of a source term as being divergence of a tensor field and thus disappearing when integrated over the whole volume.

Moreover, the total mass flux, $\uu + \jj$, is a conserved quantity, since $\uu$ is conserved and $\jj$ is divergence of a tensor field. Finally, let us discuss the possibility of rigid-body rotation. Such a rotation also implies that no compression or expansion is taking place. There are, however, two possible meanings of volume change in the weakly non-local setting, which corresponds to the two possibilities for choosing velocity, namely $\vv = \uu/\rho$ or $\bar{\vv} = \Phi_\uu$. When employing the former, we obtain $\jj=0$ from $\nabla\cdot\vv =0$, and standard fluid dynamics is recovered, which is compatible with solid-body rotation. When employing the latter definition of velocity, we obtain $\rho\bar{\vv} = \uu + \jj$. No volume change then means $\nabla\cdot \bar{\vv}$. Setting $\jj=0$ everywhere then leads to $\vv = \bar{\vv}$ and we again recover the standard fluid dynamics. Therefore, solid-body rotation is an admissible solution of evolution equations \eqref{eq.evo.rho} and \eqref{eq.evo.ui.full}.

In summary, all the criteria required in \cite{HCO-mass-flux} are satisfied in the weakly non-local setting. In particular booster and angular momentum are conserved although $\pp \neq \jj$ due to the divergence form of $\jj$. Let us now discuss physical implications of the extra term in free energy \eqref{eq.Phi}.

\section{Discussion}

The modifications of the mass flux introduced above have consequences on solutions of the fluid mechanics equations. Some of the consequences have been considered to be in disagreement  with experimental observations and  modifications of the mass flux have been therefore disallowed. In this section we discuss four objections. First is the conservation of the booster density, $\bb=\pp t - \rho\xx$, seen in \cite{KosLiu98a} as a natural requirement. The second is the set of conditions in \cite{HCO-mass-flux}. A third argument is based on  Reynolds transport theorem. Finally kinetic theory arguments are shortly discussed.

In the Kost\"{a}dt and Liu analysis \cite{KosLiu98a}, in which the momentum density is considered implicitly as four-vector, the booster density and its conservation appears naturally. However, this is not the case of our analysis, in which mass, momentum,  and energy densities are all combined into a third order four-tensor. We have introduced a definition of booster-angular momentum conservation, analogously to the special relativistic version. The booster density is Galilean invariant, as well as the condition of booster conservation, $\jj=\pp$. Also, considering the transformation properties of the pressure tensor, \re{Ptr}, we see that if booster conservation is violated, then angular momentum conservation becomes frame dependent: conservation in a particular frame does not ensure conservation in a different frame. As a conclusion we can say that the physical content and  experimental consequences of the booster conservation have  not yet been sufficiently clarified.

Let us now turn to the arguments opposing the presence of extra mass flux in balance of mass formulated in \cite{HCO-mass-flux}. They have considered four conditions:
	\begin{itemize}
		\item Galilean invariance,
		\item possibility of rigid fluid rotation,
		\item existence of locally conserved angular momentum,
		\item consistency with uniform center of mass motion.	
	\end{itemize}
We have seen here that Galilean invariance can be satisfied differently than it was assumed in \cite{HCO-mass-flux}. Transformation rules \re{rtr}-\re{qtri} are derived in a Galilean relativistic framework. Consequently, the thermodynamic force of mass diffusion is proportional to the gradient of the chemical potential, and not related OT the gradient of pressure, as it was suggested by Brenner \cite{BedEta06a}. Neither there is a need of and extra velocity, playing a central role  in the arguments of \cite{HCO-mass-flux}. The last two conditions are not independent because booster and angular momentum conservations are components of a particular four-tensor.  Therefore, the consistency with uniform center of mass motion, more properly the booster conservation, plays a central role and also offers several ways of extra mass flux with booster conservation. One of them could be the concept of internal booster, like internal angular momentum. We have given an another, related particular example in  section \ref{Ill}, where extra mass flux appears in a divergence form, which keeps the total booster conserved.  This requirement coincides with the integrability condition  of \cite{HCO-mass-flux}.

The physical background of our example is a Hamiltonian framework with kinetic energy of compression, yielding evolution equations \eqref{eq.evo.rho} and \eqref{eq.evo.ui.full}. These equations are Galilean invariant, which can be proved straightforwardly as in \cite{HCO-mass-flux}. Moreover, they obey the integrability condition required in \cite{HCO-mass-flux}, namely that the extra mass flux is a spatial gradient of a function, and they admit solid-body vortex as a solution. Finally, they fulfill angular momentum conservation, which was indicated as the most serious argument in \cite{HCO-mass-flux}.  Regarding booster conservation it is a particular example of the above mentioned combined extra mas flux and extra booster flux.

Equations \eqref{eq.evo.rho} and \eqref{eq.evo.ui.full} can be simplified in the low-Mach-number limit and when particle surface number, $\alpha$, is small. Mach number is constructed from a typical velocity, which could be shifted by an arbitrary constant by a Galilean transformation. Therefore, letting Mach number go to zero makes it impossible to require Galilean invariance because the typical velocity is fixed. That is the case of equations \eqref{eq.hydro.limit}. However, mass, momentum and angular momentum are still conserved up to order $O(\alpha)$. Extra mass flux contributes to irreversible evolution after the limit in contrast with Eq. \eqref{eq.evo.rho}. Equations \eqref{eq.hydro.limit} have also interesting physical implications. For example, they imply migration of fluid particles towards center of a Poiseuille flow.

An assumption behind the Reynolds transport theorem, see e.g. \cite{Gurtin}, is that the fluid particles are infinitely small. Indeed, only then one can divide a volume in space so that there are only particles inside or outside the volume. When the particles gain finite size, there are always some particles partially in and partially out, i.e. they are intersected by boundary of the volume. In the setting of nonlocal kinetic theory, Sec. \ref{sec.KT}, this corresponds to that when the collision kernel is strictly local in space, as in the case of classical Boltzmann equation, there is no extra mass flux while admitting non-locality in space leads to the presence of an extra mass flux. Similarly, when setting size of the fluid particles to zero in the Hamiltonian setting of Sec. \ref{Ill}, i.e. $\alpha = 0$, the extra mass flux also disappears. Reynolds transport theorem thus does not apply when considering finite size of fluid particles.

Finally we should remark that standard derivations of the governing equations fluid mechanics from kinetic theory cannot lead to extra mass flux, because the mass of particle number balances are related to the conservation of density function (see e.g. \cite{Lib90b,MulRug98b}). However, weakly nonlocal collision integrals may result in extra mass flux in the related  equations of fluid dynamics \cite{Miroslav-JSP2008}.

\section{Conclusion}

Galilean relativistic and Hamiltonian structures of fluid mechanics clarify the conditions for possibility of an extra mass flux, i.e. mass flux being different than momentum density. Conservation of total booster, the locally conserved center of mass, is fulfilled when momentum density and mass flux are equal. However, even if mass flux and momentum density are not the same, the total booster can be conserved when the extra mass flux is divergence of a tensor field. An extra mass flux was identified in Sec. \ref{Ill} which does not violate any admissibility criterion formulated in \cite{HCO-mass-flux}. The model is out of the scope of paper \cite{HCO-mass-flux}.

The physics that we are suggesting behind this extra mass flux and internal booster is an inclusion of a fine-scale motion into the setting of the classical fluid mechanics. In order to arrive at the self-diffusion contribution to the mass flux, the fine-scale motion  has to contribute to the overall kinetic energy. In the illustration worked out in this paper,  the fine-scale motion is the motion of the volume of the fluid particle characterized by $\nabla\cdot \vv$, where  $\vv$ is the field of velocity. Requirement of the Hamiltonian structure of the time-evolution equations implies that  the overall mass flux depends on $\nabla\cdot\vv$.  If  $\nabla\cdot \vv$  is let to appropriately relax in the low Mach number limit, it becomes enslaved by the gradient of mass density $\nabla\rho$ (i.e. $\nabla\cdot \vv$ becomes a function of $\nabla\rho$). This function then brings  $\nabla\rho$ into the mass flux.

We have demonstrated here that the relative balances \eqref{sm_rbal}-\eqref{se_rbal} derived from the reference frame and flow-frame independent theory are compatible with thermodynamics and Galilean transformations. The presence of the extra mass flux term is further supported by a possible physical background from weakly nonlocal Hamiltonian fluid dynamics and kinetic theory. The goal of this paper is to revive the discussion of extra mass fluxes.

\section*{Acknowledgment}
This project was supported by Natural Sciences and Engineering Research Council of Canada (NSERC) and by the grants K104260, K116197 and K116375 of the Hungarian National Research Fund. The work was supported by Czech Science Foundation (project no. 14-18938S).

\appendix
\section{Physical meaning of booster conservation}\label{sec.booster}
Consider a collection of classical point-like particles described by indexes $1,\dots,N$ and define a quantity 
\begin{equation}
\RR = \sum_{i=1}^N m_i \rr_i, 
\end{equation}
$\rr_i$ and $m_i$ being position and mass of the $i$-th particle. Vector $\RR$ expresses position of the center of mass of the particles. Time-derivative of this quantity is given by Newton's law,
\begin{equation}
\dot\RR = \sum_{i=1}^N \pp_i,
\end{equation}
where $\pp_i$ is momentum of the $i$-th particle. In the continuum approach this last equation becomes
\begin{equation}
\frac{\partial}{\partial t} \intd\rr \rho \rr = \intd\rr \uu,
\end{equation}
which can be rewritten as
\begin{equation}
\intd\rr \left(\frac{\partial \rho(t,\rr) \rr}{\partial t} - \uu(t,\rr)\right) = 0.
\end{equation}
This equation leads to the requirement that the integrand is divergence of a second order tensor field. Such requirement is equivalent (as time-derivative of momentum has the divergence form) to 
\begin{equation}
\frac{\partial b_i}{\partial t} = \partial_j B_{ij}
\end{equation}
for an unknown tensor field $\mathbf{B}$. This is the local conservation of booster.

\section{Extra mass flux of Dzyaloshinskii and Volovick}\label{sec.DV}
An extra mass flux was introduced by Dzyaloshinskii and Volovick in \cite{DV}, where the term $\nabla\cdot(D \nabla \mu)$ was added (taking quadratic dissipation function) to the right hand side of evolution equation for density. Is such an extra mass flux admissible?

Mass remains clearly conserved. Since the evolution equation for momentum is the same as in standard hydrodynamics, momentum and angular momentum are also conserved. Booster is conserved provided $D$ is constant as in the case of Eq. \eqref{Ham.b}. 

In a solid-body rotation in cylindrical coordinates the velocity is proportional the unit vector $\mathbf{e}_\phi$, perpendicular to $\mathbf{e}_r$, and it is divergence-free. Steady state balance of mass then reads
\begin{equation}
 \nabla \rho \cdot \vv + \rho \nabla\cdot \vv = D \nabla\cdot\nabla \mu.
\end{equation}
Due to radial symmetry, gradient of $\rho$ is parallel to $\mathbf{e}_r$ and thus perpendicular to $\vv$. The left hand side of this last equation thus disappears, and so the chemical potential must be constant or proportional to $1/r$. On the other hand, it follows from isothermal Gibbs-Duhem relation and from balance of forces that derivative of $\mu$ with respect to the radial direction, $\mu'$, is proportional to $\vv^2/r$. To make these two results compatible, it necessarily holds that $\vv \propto r^{-1/2}$. Such a velocity profile, however, does not fulfill the momentum balance, see e.g. p. 55 of \cite{LanLif59b}. 

In summary, all the admissibility criteria except for the possibility of solid-body rotation are fulfilled by the model of Dzyaloshinskii and Volovick. 

\section{Physical consequences of the weakly non-local free energy}\label{sec.phys.conseq}
In this section we discuss what physical implication have non-local free energy \eqref{eq.Phi} and evolution equations \eqref{eq.evo.rho} and \eqref{eq.evo.ui.full}.
\subsection{Non-dimensional form}
Firstly, we rewrite the evolution equations in a non-dimensional form in order to proceed in heuristically the same way as \cite{Lions-Masmoudi}. Speed of sound is defined as $c^2 = p_\rho$. All quantities are then expressed in non-dimensional form by 
\begin{eqnarray}
\rho \rightarrow \rho \bar{\rho}, & \partial_i \rightarrow \frac{\partial_i}{\bar{L}}, & v_i \rightarrow v_i \bar{v},  \\
& t \rightarrow \frac{\bar{L}}{\bar{v}}, & \eta_{vol} \rightarrow \eta_{vol} c^2 \frac{\bar{L}}{\bar{v}}.
\end{eqnarray}
Mach number and a new non-dimensional number expressing particle surface area
\begin{equation}
M = \frac{\bar{v}}{c} \mbox{ and } \alpha = \frac{\sigma}{\bar{L}^2}
\end{equation}
will appear in the evolution equations. The latter number expresses how small surface of the particles is. Evolution equations \eqref{eq.evo.rho} and \eqref{eq.evo.ui.full} then become
\begin{subequations}\label{eq.evo.lM}
	\begin{eqnarray}
	\label{eq.evo.rho.lM}\frac{\partial \rho}{\partial t} &=& -\partial_i (\rho v_i) + \alpha \partial_i\partial_i(\rho \partial_j v_j)\\
	\label{eq.evo.ui.lM}\frac{\partial u_i}{\partial t} &=& -\frac{1}{M^2}\rho \partial_i \varphi_\rho - \partial_j \left(\frac{u_i u_j}{\rho}\right) \nonumber\\
	&&- \alpha\partial_j \left(\partial_i v_j \rho \partial_k v_k - v_i \partial_j( \rho \partial_k v_k)\right)\nonumber\\
	&& +\frac{1}{M^2}\partial_i\left(\eta_{vol}\rho \partial_j\left(\frac{u_j}{\rho}-\alpha\frac{1}{\rho}\partial_j \left(\rho\partial_k \frac{u_k}{\rho}\right)\right)  \right).
	\end{eqnarray}
\end{subequations}

\subsection{Low Mach number and surface are limit}
In the limit of low Mach number, $M\rightarrow 0$, and when the surface area is small, $\alpha<<1$, we obtain the condition
\begin{subequations}
	\begin{equation}
	\rho \partial_i \varphi_\rho = \partial_i\left(\eta_{vol}\rho \partial_j\left(\frac{u_j}{\rho}-\alpha\frac{1}{\rho}\partial_j \left(\rho\partial_k \frac{u_k}{\rho}\right)\right)  \right), 
	\end{equation}
	or 
	\begin{equation}\label{eq.cond2}
	\partial_i \left(\rho \partial_k v_k \right) = \frac{1}{\eta_{vol}}\left(\rho \partial_i \varphi_\rho\right) + \alpha \partial_i\left(\rho\partial_j \left(\rho^{-1}\partial_j\left(\rho \partial_k v_k\right)\right)\right).
	\end{equation}
	Since the first term on the right hand side of this last equation can be rewritten as gradient of pressure, the condition can be also rewritten as
	\begin{eqnarray}\label{eq.cond3}
	\rho \partial_k v_k  &=&  \frac{p-C}{\eta_{vol}} + \alpha \rho\partial_j \left(\rho^{-1}\partial_j\left(\rho \partial_k v_k\right)\right)\nonumber\\
	&=&\frac{p-C}{\eta_{vol}} +\frac{\alpha\rho}{\eta_{vol}}\partial_j\left(\frac{1}{\rho}\partial_j p\right) + O(\alpha^2)
	\end{eqnarray}
	where $C$ is a constant. 
\end{subequations}

Plugging condition \eqref{eq.cond2} into evolution equation \eqref{eq.evo.rho.lM} and conditions \eqref{eq.cond2} and \eqref{eq.cond3} into Eq. \eqref{eq.evo.ui.lM} and dropping all terms of order $O(\alpha^2)$ yields
\begin{subequations}\label{eq.hydro.limit}
	\begin{eqnarray}
	\frac{\partial \rho}{\partial t} &=& -\partial_i u_i -  \partial_i j_i\\
	\label{eq.evo.ui.final}\frac{\partial u_i}{\partial t} &=& -\partial_j \left(\frac{u_i u_j}{\rho}\right)\nonumber\\
	&&-\frac{\alpha }{\eta_{vol}}\partial_j\left(\partial_i v_j (p-C) - v_i \partial_j p\right)\nonumber\\
	&=& -\partial_i \pi - \partial_j\left(\frac{u_i u_j}{\rho}\right)\nonumber\\
	&&+\frac{\alpha }{\eta_{vol}}\left(\mathbf{\omega} \times \nabla p\right)_i + \frac{\alpha}{\eta_{vol}} v_i \partial_j\partial_j p
	\end{eqnarray}
\end{subequations}
where the extra mass flux was identified as 
\begin{equation}
\jj = -D \nabla \varphi_\rho \mbox{ with } D = \frac{\alpha\rho }{\eta_{vol}},
\end{equation}
a normalized pressure $\pi$ is
\begin{equation}
\pi = \frac{\alpha (p-C)^2}{2\eta^2_{vol}}
\end{equation}
and $\mathbf{\omega}$ is vorticity,
\begin{equation}
\omega_i = \eps_{ijk}\partial_j v_k.
\end{equation}
In this low Mach number and low surface area limit, the extra mass flux obviously produces irreversible evolution and leads to homogenization of density, since chemical potential can be identified as $\mu = \varphi_\rho$. 

The term with vorticity is an irreversible term that could be responsible for particle migration. Indeed, consider a Poiseuille 2D flow in an infinite rectangular channel to the right, where pressure decreases in the $x$-direction, $y$ is perpendicular to the flow and $z$-direction positive so that the basis is right-oriented, i.e. points upwards. Then in the upper half of the channel $\omega$ is positive in the $z$-direction, $\nabla p$ is negative in the $x$-direction and the vorticity term in Eq. \eqref{eq.evo.ui.final} is negative in the $y$-direction, i.e. produces force to pointing to the center of the channel. See Sec. \ref{sec.migration} for more details.

The last term in Eq. \eqref{eq.evo.ui.final} is also irreversible (indeed, it is odd with respect to time-reversal), and it reduces the velocity if $\Delta p$ is negative. 

Mass is clearly conserved in equations \eqref{eq.hydro.limit}. What about total momentum? From Eq. \eqref{eq.evo.ui.final} it follows that
\begin{eqnarray}\label{eq.ui.cons}
\frac{\partial}{\partial t}\intd\rr u_i &=& \frac{\alpha }{\eta_{vol}}\intd\rr \left(\eps_{ijk} \eps_{jmn} \partial_m v_n \partial_k p  + v_i \partial_k\partial_k p\right)\nonumber\\
&=&\frac{\alpha}{\eta_{vol}}\intd\rr\left(\partial_k v_i \partial_k p - \partial_i v_k \partial_k p\right) - \partial_k v_i \partial_k p \nonumber\\
&=&-\frac{\alpha }{\eta_{vol}}\intd\rr \partial_k v_k \partial_i p = \frac{\alpha }{\eta}\intd\rr \left(\frac{p-C}{\eta\rho} + O(\alpha)\right)\partial_i p
\end{eqnarray}
Pressure $p$ is a function of density. Therefore, there exists a function $f(\rho)$ such that
\begin{equation}
f'(\rho) = \frac{p-C}{\rho}\frac{\partial p}{\partial \rho}.
\end{equation}
When neglecting all terms of order $O(\alpha^2)$ and higher, Eq. \eqref{eq.ui.cons} becomes
\begin{equation}
\frac{\partial}{\partial t}\intd\rr u_i = \frac{\alpha }{\eta^2_{vol}}\intd\rr \partial_i f = 0,
\end{equation}
and conservation of momentum thus holds also after the limit $M\rightarrow 0$ and $\alpha << 1$.

And what about angular momentum conservation? Firstly, the symmetric part of the stress in Eq. \eqref{eq.evo.ui.final} can not violate the conservation. Let us therefore focus on the apparently non-symmetric part, which contributes to evolution of total angular momentum as follows:
\begin{eqnarray}\label{eq.a.cons}
\frac{\partial}{\partial t}\intd\rr \eps_{lki}r^k u_i &=& \frac{\alpha }{\eta_{vol}} \intd\rr \left(\eps_{lki}r^k \eps_{imn} \omega_m \partial_n p +\eps_{lki}r^k v_i \partial_j\partial_j p\right)\nonumber\\
&=& \frac{\alpha}{\eta_{vol}} \intd\rr r^k(\omega_l \partial_k p - \omega_k \partial_l p)\nonumber\\
&&-\frac{\alpha}{\eta_{vol}}\intd\rr (\underbrace{\eps_{lji}v_i \partial_j p}_{=-\omega_l p}+ \eps_{lki} r^k \partial_j v_i \partial_j p)\nonumber\\
&=&\frac{\alpha}{\eta_{vol}} \intd\rr \left(r^k(\omega_l \partial_k p - \omega_k \partial_l p )+ \omega_l p\right)\nonumber\\
&&-\frac{\alpha}{\eta_{vol}} \intd\rr \eps_{lki}r^k (\partial_i v_j + \eps_{jim}\omega_m)\partial_j p\nonumber\\
&=&\frac{\alpha}{\eta_{vol}} \intd\rr \left(r^k(\omega_l \partial_k p - \omega_k \partial_l p )+ \omega_l p\right)\nonumber\\
&& +\frac{\alpha}{\eta_{vol}} \intd\rr \left(r^k(\omega_k \partial_l p -\omega_l \partial_k p) - \eps_{lki}r^k \partial_i v_j \partial_j p\right)\nonumber\\
&=&\frac{\alpha}{\eta_{vol}} \intd\rr (\omega_l p + \underbrace{\eps_{lji}\partial_i v_j p}_{=-\omega_l p} + \eps_{lki}r^k \partial_i\partial_j v_j p)\nonumber\\
&=& \frac{\alpha}{\eta_{vol}}\intd\rr \eps_{lki} r^k \partial_i \left(\frac{p-C}{\eta\rho}\right)p = -\frac{\alpha}{\eta_{vol}}\intd\rr \eps_{lki} r^k \frac{p-C}{\eta\rho}\partial_i p\nonumber\\
&=& -\frac{\alpha}{\eta_{vol}}\intd\rr \eps_{lki}r^k \partial_i f = \frac{\alpha}{\eta_{vol}}\intd\rr \eps_{lii} f = 0.
\end{eqnarray}
where terms of order $O(\alpha^2)$ were neglected. Angular momentum is thus also conserved up to order $O(\alpha)$ after the limit.

\subsection{Particle migration}\label{sec.migration}
The extra force term $\omega\times\nabla p$ in Eq. \eqref{eq.evo.ui.final} causes migration of particles towards center of a channel (or tube) exhibiting Poiseuille flow. Let us now compare qualitatively the effect of the term with experimental results concerning particle migration. Assuming velocity profile
\begin{equation}
v_x = - \frac{1}{\mu} \partial_x p y
\end{equation}
where $\mu$ is dynamic shear viscosity of the fluid, the extra force term becomes
\begin{equation}
-\frac{\alpha}{\eta_{vol}\mu}(\partial_x p)^2 y \mathbf{e}_y, 
\end{equation}
which corresponds to an effective potential 
\begin{equation}
\varphi_m = \frac{\alpha}{2\eta_{vol}\mu} (\partial_x p)^2 y^2.
\end{equation}

\begin{figure}[ht]
	\begin{center}
		\includegraphics[width=0.5\textwidth]{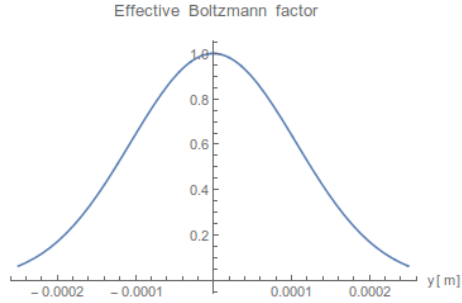}
		\caption{\label{fig.migration}Effective Boltzmann factor \eqref{eq.BF} for data from \cite{Weeks}.}
	\end{center}
\end{figure}

In physical units the potential becomes
\begin{equation}
\varphi_m = \frac{\sigma}{2\eta_{vol}\mu} (\partial_x p)^2 y^2, 
\end{equation}
which has units $J/m^3$. Energy of a particle with volume $V$ at $y$ is $V \varphi_m(y)$. The corresponding Boltzmann factor
\begin{equation}\label{eq.BF}
f(y) \propto \exp\left(- \frac{V \varphi_m(y)}{kT}\right)
\end{equation}
then indicates that there are more particles near center of the channel. Taking data from \cite{Weeks}, $L = 5\mu m$, $\rho = 1.23 g cm^{-3}$, spherical particle diameter $2.3\mu m$, shear viscosity of bromocyclohexane $\mu = 2.8 \cdot 10^{-3} Pa\,s$, bulk viscosity estimated as that of cyclohexanone \cite{Dukhin}, i.e. $\eta_{vol} = 9\cdot 10^{-3} Pa\,s/\rho$, maximum velocity in Poiseuille profile $v_{max} = 2.95 \mu m\,s^{-1}$, which gives the pressure gradient as $\partial_x p = 8\mu v_{max}/L^2$, temperature $T = (273.15+22)K$, the Boltzmann factor can be plotted as in Fig. \ref{fig.migration}.

The effective Boltzmann factor only provides a crude approximation of concentration profile of particles as also the effect of the hydrodynamic field should be taken into account. Moreover, the volume (or bulk) viscosity was only estimated. However, the profile clearly shows migration towards the center of the channel.

In summary, regarding kinetic energy of volume expansion leads to the presence of extra terms in the reversible evolution of hydrodynamics. After volumes of fluid particles relax to a stationary, the coupling between momentum and density provided by the Poisson bracket becomes irreversible, and the irreversible mass flux naturally appears. Besides  irreversible mass flux, also irreversible stress appears that leads for example to particle migration towards center of a channel in a Poiseuille flow.

%\bibliographystyle{ieeetr}
%\bibliography{library}

\end{document}